\documentclass[twocolumn,apsrev4-2,aps,physrev,superscriptaddress,floats,amsmath,amssymb,floatfix,longbibliography,]{revtex4-2}

\usepackage{color}
\usepackage{amsmath}
\usepackage{graphicx}
\usepackage[pdf]{pstricks}
\usepackage{pdftexcmds}
\usepackage{amssymb}
\usepackage{esint}
\usepackage{comment}
\usepackage{physics}
\usepackage{bm}
\usepackage{lipsum}    
\usepackage[utf8]{inputenc}

\def \bea {\begin{eqnarray}}
\def \eea {\end{eqnarray}}

\allowdisplaybreaks

\begin{document}

\title{Unsupervised machine learning for the detection of exotic phases in   skyrmion phase diagrams}

\author{F. A. G\'omez Albarrac\'in}
\email{albarrac@fisica.unlp.edu.ar}
\affiliation{Instituto de Física de Líquidos y Sistemas Biológicos (IFLYSIB), UNLP-CONICET, Facultad de Ciencias Exactas, 1900, La Plata, Argentina}
\affiliation{Departamento de Ciencias Básicas, Facultad de Ingeniería, Universidad Nacional de La Plata, 1900, La Plata, Argentina}

%\date{\today}

\begin{abstract}
Undoubtedly, machine learning (ML) techniques are being increasingly applied to a wide range of situations in the field of condensed matter. Amongst these techniques, unsupervised techniques are especially atractive, since they imply the possibility of extracting information from the data without previous labeling. In this work, we resort to the technique known as ``anomaly detection'' to explore potential exotic phases in skyrmion phase diagrams, using two different algorithms: Principal Component Analysis (PCA) and a Convolutional Autoencoder (CAE). First, we train these algorithms with  an artificial dataset of skyrmion lattices constructed from an analytical parametrization, for different magnetizations, skyrmion lattice orientations, and skyrmion radii. We apply the trained algorithms to a set of snapshots obtained from Monte Carlo simulations for three ferromagnetic skyrmion models: two including in-plane Dzyaloshinskii-Moriya interaction (DMI)  in the triangular and kagome lattices, and one with an additional out-of-plane DMI in the kagome lattice. Then, we compare the root mean square error (RMSE) and the binary cross entropy (BCE) between the input and output snapshots as a function of the external magnetic field and temperature. We find that the RMSE error and its variance in the CAE case may be useful to not only detect exotic low temperature phases, but also to differentiate between the characteristic low temperature orderings of a skyrmion phase diagram (helical, skyrmions and ferromagetic order). Moreover, we apply the skyrmion trained CAE to two antiferromagnetic models in the triangular lattice, one that gives rise to antiferromagnetic skyrmions, and the pure exchange antiferromagnetic case, finding that the behaviour of the RMSE is still an indicator of different ordering. Finally, we explore the portability of the technique with the same CAE applying it to a ferromagnetic skyrmion model in the square lattice.
%Despite the predictably larger RMSE, we find that, even in these cases, the RMSE is also an indicator of different orderings and the emergence of particular features, such as the well-known  pseudo-plateau  in the pure exchange case.

\end{abstract}

\maketitle

%%%%%%%%%%%%%%%%%%%%%

\section{Introduction}

 Machine Learning (ML) has been increasingly incorporated into a wide variety of technological and scientific research in the last few years. In condensed matter physics, some of the first studies using ML include the representation of quantum states \cite{Troyer}, the identification of phases of matter \cite{Carrasquilla2017}, inverse design in photonics \cite{Sheverdin2020} and the study of phase transitions \cite{Rem,Wang2016,vanNieuwenburg2017}. In particular, ML has been used to explore different aspects of magnetic skyrmions, which are topological textures \cite{Bogdanov1989,Bogdanov1994,Rossler2006,Gobel2021,Tokura2021} with potential technological applications, especially in memory storage devices \cite{Nagaosa2013,Fert2013,Fert2017}. Feed forward and convolutional neural network have been used to classify different magnetic phases and predict features in simple skyrmion models using simulations data \cite{Iakovlev2018,Singh2019,SalcedoGallo2020,Kawaguchi2021,MLSkyAlbarracin2022,Araz2022,Feng2024,Mazurenko2023}. Moreover, ML algorithms have been applied also  to experimental data to explore skyrmion phases \cite{Wang2021,Matthies2022,Labrie2024}.

In general, ML techniques could be devided in three large types of techniques  \cite{Bishop, Geron}: supervised, unsupervised and reinforcement learning, the latter one being the case where the algorithm ``learns'' solely from the data. In supervised ML, a previous labeling and knowledge of the data is needed before applying the algorithms to an unknown dataset, and is widely used in classification and regression tasks. On the other hand, no data labeling is required in unsupervised ML,  and some of these techniques may be used for example to group similar instances of  the data (clustering) or detect atypical instances (anomaly detection). A variety of this type of unsupervised (or semi-supervised) methods has been used in condensed matter physics, from Principal Component Analysis (PCA) in Ising systems \cite{Wang2016} to support vector machines with tensorial kernels applied to models in different types of topology \cite{Pollet2019op1,Pollet2019op2,Pollet2019piro,Pollet2021JPCM,Pollet2021PRE,Pollet2021Kitaev1,Pollet2021Kitaev2,Pollet2023},  the use of autoencoders  to study neutron scattering data in spin ice systems \cite{Samarakoon2020,Samarakoon2022},  the identification of superconducting phases \cite{Tibaldi2023}  and to explore anomaly detection in simple frustrated models \cite{Acevedo}.

In this work, we resort to anomaly detection to explore possible unusual phases in skyrmion systems. In order to this, we train two techniques (PCA and Convolutional Autoencoders, CAE) to process an analitically generated (``artificial'') skyrmion crystal dataset. Then, we use the trained algorithms on spin configurations obtained through Monte Carlo simulations from different ferromagnetic skyrmion models at different temperatures under magnetic fields, and calculate the root mean squared error (RMSE) between the input and output snapshot. We find that both the mean value and the deviation of the RMSE  from the data reconstructed with the CAE may be used to distinguish between three typical phases found in skyrmion systems (helices, skyrmion crystals and ferromagnetic), and, most importantly, to detect possible exotic orderings, such as bimeron glasses at higher magnetic fields \cite{CSLSky2023}. We also apply the trained CAE to snapshots from two antiferromagnetic models, and explore what information  may be obtained from the RMSE.

The manuscript is organised in the following way. In Sec. II we go over relevant characteristics from magnetic skyrmions, and describe the ferromagnetic skyrmion models chosen in this work. The ML scheme and details on  the dataset and the algorithms  are given in Sec. III. We discuss our results in Sec. IV, where we also consider other antiferromagnetic models,  and present conclusions and perspectives  in Sec. V.

\begin{table*}[th!]
\centering
\begin{tabular}{|c|c|c|c|}
\hline 
& & & \\
& \large{Triangular (TFDM$^{xy}$)} & \large{Kagome (KFDM$^{xy}$)} & \large{Kagome + $D^z$ (KFDM$^z$)}
 \\
& & & \\
\hline 
& & & \\
\rotatebox{90}{\large{model}} & \begin{minipage}{.3\textwidth}
      \includegraphics[width=\linewidth]{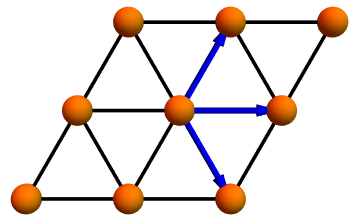}
    \end{minipage} & \begin{minipage}{.3\textwidth}
      \includegraphics[width=0.9\linewidth]{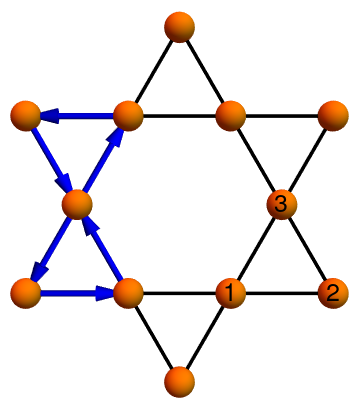}
    \end{minipage} & \begin{minipage}{.3\textwidth}
      \includegraphics[width=0.9\linewidth]{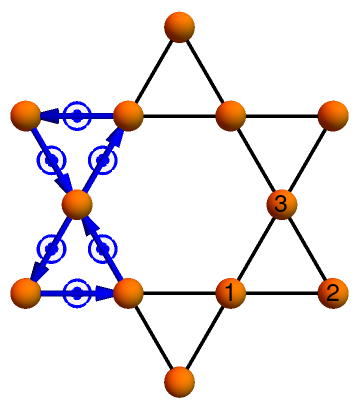}
    \end{minipage}\\
\hline 
& & & \\
\large{ $J/|J|$ } & \large{-1} &  \large{-1} & \large{-1} \\
& & & \\
\hline
& & & \\
\large{$D^{xy}/|J|$} & \large{1.5} &  \large{0.5} & \large{0.5} \\
& & & \\
\hline 
& & & \\
\large{$D^z/|J|$} & \large{0} & \large{0} & \large{$\sqrt{3}$} \\
& & & \\
\hline 
\end{tabular}
\caption{\label{tab:latt} Three ferromagnetic spin models used in this work, with  nearest neighbor exchange ($J$) and Dzyaloshinskii Moriya interactions (DMI). The blue arrows indicate the in-plane DMI (with strength $D^{xy}$) and the circles indicate the direction of the out-of plane DMI ($D^z$). The rows show a sketch of the lattice, the value of the exchange interaction (which in this case is always  $J/|J|=-1$ , ferromagnetic), and of the in-plane and out-of-plane DMI strengths.}
\end{table*}

\section{Brief overview on skyrmions and spin models} \label{sec:model}

Magnetic skyrmions are an arrangement of spins characterised by the topological charge, defined as $Q=\frac{1}{4\pi}\int d^2r \vec{S}\cdot(\partial_x\vec{S}\times\partial_y\vec{S})$;  where $\vec{S}$ are the spin unit vectors. Skyrmions are especially suitable for promising technological applications due to their robustness, stability and easy manipulation. A large number of materials and models have been shown to host skyrmions \cite{Muhlbauer2009,Yu2010} and skyrmion like textures (such as antiferromagnetic skyrmions \cite{Rosales2015,Gao2020}, magnetic bubbles, antiskyrmions \cite{Dohi2019,Mukherjee2022,Mohylna2022}, merons \cite{Lin2015}, etc). A skyrmion crystal or lattice in a bidimensional system is a periodic arrangement of skyrmions and may simply be parametrised as the superposition of three non-coplanar helices \cite{Osorio2017,Gobel2021}; the spin parametrization $s_{SkX}$ at position $\mathbf{r}$ is given by :

\begin{eqnarray} \label{eq:skxpara}
s_{SkX}(\mathbf{r}) = \frac{1}{s}\left(\sum_{\mu=1}^3\sin\left(\mathbf{q}_{\mu}\cdot\mathbf{r} + \theta_{\mu}\right)\mathbf{e}_{xy,\mu}\right. + \\ \nonumber
\left.\left[ \sum_{\mu=1}^3\cos\left(\mathbf{q}_{\mu}\cdot\mathbf{r} + \theta_{\mu}\right)\mathbf{e}_{z,\mu} + m_0 \right] \right)
\end{eqnarray}

\noindent where $s$ normalizes $|s_{SkX}(\mathbf{r})|$  to 1,  $m_0$ is the homogenous contribution to the magnetization in
the $z$ direction, arbitrary unit vectors in the $xy$ plane $\mathbf{e}_{xy,\mu}$ satisfy  $\sum\mathbf{e}_{xy,\mu}=0$, and $\theta_{\mu}$ are the phase factors that fullfill $\cos\left(\sum_{\mu} \theta_{\mu}\right)=-1$. The
$\mathbf{q}_{\mu}$ vectors are the three ordering vectors of the helices,  that lie in the same plane  at 120$^{\circ}$, with norm $2\pi/r_0$, $r_0$ being the skyrmion radius. 

\begin{figure}[h!]
    \centering
    \includegraphics[width=1.0\columnwidth]{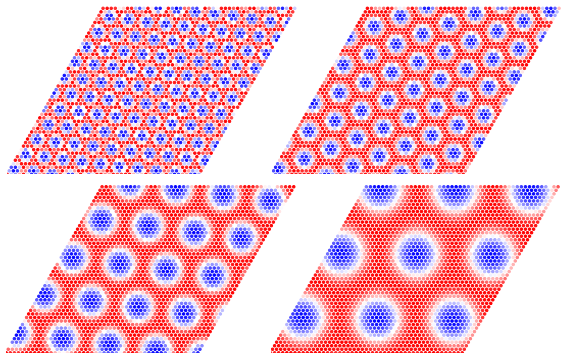}
    \caption{ Example of skyrmion lattice snapshots constructed with the analytical solution in Eq.~(\ref{eq:skxpara}) used for training the ML algorithms.}
    \label{fig:train}
\end{figure}

%The structure factor, which is the Fourier transform of the spin-spin correlations, of this type of arrangement gives six bright peaks, corresponding to the three inequivalent $q$ ordering vectors. Thus, a ``triple-q'' pattern in reciprocal space is a strong suggestion that a skyrmion lattice is present, as measured first in bulk MnSi \cite{Muhlbauer2009}. 
 A relevant parameter that may be calculated in a discrete lattice  of $N$ spins is the scalar chirality, the sum of the triple product of neighbouring spins ($i,j,k$) forming a triangle, defined as:

\begin{equation} \label{eq:chi}
\chi=\sum_i^N  \vec{S}_i \cdot \left(\vec{S}_j\cross\vec{S}_k\right)
\end{equation}

This quantity, with a factor of $\frac{1}{8\pi}$, is the discrete version of the topological charge, which is  $|Q|=1$ for skyrmions \cite{Gobel2021},  and is related to the skyrmion number \cite{Onoda2004}. Therefore, for a given arrangement of spins, a non zero value of the chirality may be an indicator of skyrmions, regardless of whether they are organized periodically or not.% To illustrate the above discussion, in Fig. 1 we show a skyrmion crystal constructed on a triangular lattice following Eq. (1), setting MM and QQQ, its corresponding structure factor and the value of the scalar chirality, XXX, which, although close, does not give the exact number of skyrmions due to the discretisation.

As was introduced in \cite{Yu2010}, a possible simple model to realise these periodic skyrmion arrangements in a bidimensional lattice is one where ferromagnetic nearest neighbors coupling competes with antisymmetric in-plane Dzyaloshinskii Moriya  interactions (DMI) \cite{dzyaloshinsky1958,moriya1960} under an external magnetic field:

\begin{equation} \label{eq:H}
\mathcal{H}=J\sum_{\langle i,j \rangle} \vec{S}_i\cdot\vec{S}_j + \sum_{\langle i,j \rangle} \vec{D}\cdot (\vec{S}_i\times\vec{S}_j) - \vec{B}\sum_i \vec{S}_i
\end{equation}

\noindent  here the magnetic moment or spin $\vec{S}_i$ is a  three dimensional unit vector  $|S_i|=1$  (Heisenberg spins) at site $i$,  $J<0$ is the ferromagnetic exchange coupling, $\vec{D}$ the in-plane DMI along the bonds of the chosen lattice ($\vec{D}=\vec{D}^{xy}$, with absolute value $D^{xy}$), and $\vec{B}=B\breve{z}$ an external magnetic field perpendicular to the lattice plane.  In this model, the skyrmion radius is related to the $D^{xy}/|J|$ ratio, a larger ratio implies smaller skyrmions. 

\begin{figure}[h!]
\includegraphics[width=1.0\columnwidth]{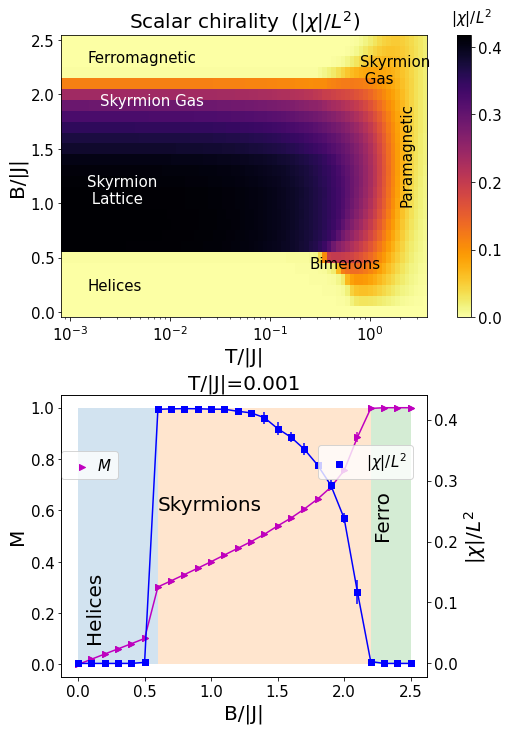} \caption{  (Top) Phase diagram for the typical skyrmion Hamiltonian form Eq.~(\ref{eq:H}) constructed using the scalar chirality (Eq.(\ref{eq:chi})) as a function of temperature and magnetic field, obtained through Monte Carlo simulations for a triangular lattice with  $48\times 48$ sites. (Bottom) Magnetisation and scalar chirality density as a function of the magnetic field at the lowest simulated temperature. }
\label{fig:skyPD}
\end{figure}

We present a typical phase diagram for the  model in Eq.(\ref{eq:H}) , in Fig.~\ref{fig:skyPD}, top panel, constructed with the scalar chirality  as a function of temperature $T$ and magnetic field $B$, obtained from Monte Carlo simulations in a triangular lattice with  $D^{xy}/|J|=1.5$ and $N=L^2$ sites ($L=48$).  At low temperatures and no external field a  helical order is found. Under an increasing external magnetic field, a skyrmion lattice is stabilized, and then at higher fields all spins are aligned with the field giving rise to a ferromagnetic or field polarized phase. Intermediate phases are enhanced with temperature:  a bimeron phase at lower fields, between helices and the skyrmion lattice, as helices are broken and distorted into bimerons; and a skyrmion gas at higher magnetizations, between the skyrmion lattice and the ferromagnetic ordering, where the skyrmion lattice order is lost and a smaller number of  skyrmions are found in a non-periodic arrangement as the system is further magnetized \cite{Yu2010, Ezawa2011}.  In this work,  the bimerons look as ``elongated skyrmions'', and are actually a pair of merons (topological structures with $|Q|=1/2$), see for example \cite{Ezawa2011,Leonov2024}, as opposed to meron - antimeron pairs  as defined in \cite{Tetriakov2019}.  Typical configurations of these phases are shown in Fig.~\ref{fig:snaps}.

In the bottom panel of Fig.~\ref{fig:skyPD} we focus on the low temperature phases, and compare the absolute value of the scalar chirality density $|\chi|/L^2$ with the magnetisation $ M =\langle \frac{1}{N} \sum_i S_i^z \rangle$. We see that there are indeed three main phases: at low field there are helices with no chirality; at larger fields the chirality dramatically increases (and the magnetisation shows a sharp change), indicating the presence of  skyrmions. As the field is increased, the number of skyrmions is reduced, reflected in a smaller $|\chi|/L^2$, and then the system goes into the ferromagnetic phase ($ M  = 1$), where the chirality is null. The background colors are a guide to the eye and indicate three type of phases: a helical phase, a skyrmion phase (which includes skyrmion lattices and a small skyrmion gas region) and a ferromagnetic or field polarized order.

The inclusion of additional interactions may imply deviations from this simple scheme. For example, it has been shown that an increasing additional in plane site anistropy may take the skyrmion lattice to an arrangement of merons \cite{Lin2015}, and more complex phase diagrams may also be found in models with frustrating exchange interactions and no DMI \cite{Okubo2012,Leonov2015}.

 Since  we aim to explore techniques that may detect exotic phases, we focus on three models. First, we take the Hamiltonian from Eq.~(\ref{eq:H}) in two triangular based lattices: triangular and kagome, which may itself be divided into three triangular lattices formed by the three types of spins in each elementary plaquette. Then, as a third model we take the Hamiltonian from Eq.~(\ref{eq:H}) in the kagome lattice and include an out-of-plane DMI ($\vec{D}=\vec{D}^{xy} + \vec{D}^z$) at the specific value  $D^z/|J|=\sqrt{3}$. 
 It has been shown that in the ferromagnetic kagome lattice with out-of-plane DMI  $D^z/|J|=\sqrt{3}$ ($D^{xy}/|J|=0, B/|J|=0$) a chiral spin liquid emerges, where the in-plane spin components retain the degeneracy and correlations of an algebraic classical spin liquid, and the system has a net magnetization \cite{Essafi2016}. The combination of this out-of-plane  $D^z/|J|$ and in plane $D^{xy}/|J|$ gives rise to several exotic phenomena,  most notably a stabilization of a skyrmion gas  with well defined skyrmion structures at higher temperatures, which precedes the formation of a skyrmion lattice,  and  a ``bimeron glass'' phase at higher magnetic fields and lower temperatures, between the skyrmion crystal and the fully polarized phase \cite{CSLSky2023,CSLSky2024}, where the system retains chiral spin liquid behavior in the $xy$ plane. This fully polarized phase extends at lower magnetic fields and intermediate temperatures, and allows the control of the number of skyrmions with temperature, going from a skyrmion gas  to a skyrmion lattice  as the temperature is lowered, which is not the typical way that skyrmion lattices are stabilized   in two dimensional systems \cite{nishikawa2019,huang2020,balavz2021}. 
 Here, we will show that it may be possible to detect these phases  using unsupervised machine learning techniques, without inspection of the snapshots.
 
In Table \ref{tab:latt} we list these three different spin models. We label the one in the triangular lattice with ferromagnetic exchange coupling and in-plane DMI along the bonds TFDM$^{xy}$. In analogy, the same model but in the kagome lattice is named KFDM$^{xy}$. Finally, the KFDM$^z$ spin model is as the KFDM$^{xy}$ with additional out-of-plane DMI, $D^z$.   Unless otherwise indicated, we will refer as TFDM$^{xy}$ as the model in the triangular lattice with  $D^{xy}/|J|=1.5$, where the radius of the resulting skyrmions be similar to that of the KF spin models. Nonetheless, we will also test our analysis  comparing results for other values of the DMI in the triangular lattice. %Other values of the DMI interaction indicated in the table ($D^{xy}=0.5, 1$) where 

%Notice that $J$ is fixed to $-1$ in all three spin models, but the $D^{xy}$ is larger for the triangular lattice. This is simply so that the radius of the resulting skyrmions be similar to that of the KF spin models.

\section{Machine Learning scheme and dataset}

In this work, we resort to the technique known as \textit{anomaly detection} to obtain information from skyrmion models, such as if it is possible to distinguish different phases and/or to detect exotic ones. The anomaly detection technique is usually described to detect events such as fraud or spam \cite{Patel}, or new trends in a time series \cite{Geron}. Simply put, the main idea is that the large amount of ``regular'' data may have certain general characteristics, that the algorithm must ``learn'', and that points that deviate from this, ``outliers'', may be worth looking into. Here, we will consider skyrmion lattices as the ``regular'' data, and analyze what possible information may be obtained from anomalies or deviations.

 In order to do this, we will train two types of algorithms, Principal Component Analysis (PCA) and a Convolutional AutoEntonder (CAE), to reconstruct a set of different skyrmion lattices. Then, we  apply the trained algorithm to spin configurations obtained from Monte Carlo simulations of different skyrmion hosting models, and calculate the root mean square error (RMSE) between the reconstructed configuration and the original one. Since the algorithms are trained for skyrmion crystals, a bigger error would imply that the  spin arrangement is further from a skyrmion lattice. 
 %We illustrate this scheme in Fig. XXX. 

  The RMSE is calculated as:
 
\begin{equation}  \label{eq:rmse}
RMSE=\sqrt {\frac{1}{N}\sum_i^N\left(S^z_i - \tilde{S}^z_i\right)^2}
\end{equation} 

\noindent where $S^z_i$ is the component along the magnetic field at site $i$ of the input data, and  $ \tilde{S}^z_i$ is that of the output (or reconstructed) data.  Here, as was done before \cite{Iakovlev2018,MLSkyAlbarracin2022}, we only consider the $S^z$ projection of the spins (renormalized to take values between 0 and 1), which reduces significantly the volume of data. This may also be interpreted as taking the spin configurations as images, and thus this work may be potentially applied to experimental images obtained with spin-polarized scanning tunneling microscopy techniques \cite{Yu2010,Bergmann2014,Hirschberger2019,Yasui2020}.

For the CAE outputs, we have also calculated de binary cross entropy (BCE) \cite{Bishop}, defined as:

\begin{equation}  \label{eq:bce}
BCE=-\frac{1}{N}\sum_i^N \left(S^z_i\log(\tilde{S}^z_i) +  (1 - S^z_i)\log(1-\tilde{S}^z_i)\right)
\end{equation}

The BCE is a quantity usually used as a loss variable to train machine learning algorithms, such as autoencoders. In the results section, we will show that, for the chosen models, in general resorting to the BCE to use the anomaly detection approach does not provide as much information as the use of the RMSE.

  In summary, we consider the following scheme:
 
 \begin{enumerate}
 \item{We train two types of algorithms, PCA and CAE, with a set of different skyrmion lattices. }
 \item{We  apply the trained algorithm to spin configurations obtained from Monte Carlo simulations of different skyrmion hosting models, and calculate the root mean square error (RMSE) between the decoded configuration and the original one. }
 \item{We study the behaviour of the RMSE and BCE for different parameters and comparing models, to evaluate whether we may obtain information on the physics of each model, and explore the possibility of using  the anomaly  technique comparing the errors (RMSE and BCE)  to find exotic phases in skyrmion-hosting systems.}
 \end{enumerate}

Below, we describe in detail the dataset used to train the algorithms, the dataset obtained from the simulations, and give details on the ML models used.

\subsection{Training dataset}

We construct ``artificial'' skyrmions cristals in triangular lattices with  $N=L^2$ ($L=48$)  sites, following the parametrization presented in Eq.~(\ref{eq:skxpara}). We take 7 possible values for the skyrmion radii $r_0$ and 10 values of the parameter $m_0$, which after normalization changes the magnetization of the lattice. Since in our analysis we will be using snapshots from MC simulations as images, in this constructed dataset we do not require periodic boundary conditions, but we must take into account rotations and translations of the parametrization pattern, which we do considering  8 possible rotations and  3 different translations in the position along the 3 directions of the nearest neighbors bonds of the triangular lattice. Therefore, we have a training and validation dataset of $7 \times  10 \times 8 \times 3 \times 3 = 5040$ different skyrmion crystals in triangular lattices.  A few examples of the training snapshots with different skyrmion sizes are those illustrating the parametrisation in Fig.~\ref{fig:train}.

\subsection{Monte Carlo simulations dataset}

We performed Monte Carlo Metropolis-Hastings simulations, combined with microcanonical updated (overrelaxation) on the three spin models described in the previous section: (1) TFDM$^{xy}$: a ferromagnetic exchange and in plane DMI model in a  $L^2, (L=48)$ triangular lattice under a magnetic field, fixing  $D^{xy}/|J|=1.5$  (2) KFDM$^{xy}$: a similar model in a  $3L^2$  kagome lattice, for  $D^{xy}/|J|=0.5$ (3) KFDM$^z$: the model presented in Eq.~(\ref{eq:H}) including an out-of-plane Dzyaloshinskii-Moriya interaction  $D^z/|J|=\sqrt{3}$. Simulations were done using periodic boundary conditions, and lowering the temperature from high temperature in 80 steps using the annealing technique. Up to 10 copies with independent seeds were done for each model and magnetic field. To apply the algorithms to the kagome snapshots, we chose one of the three sublattices (each one corresponding to the three types of sites in the kagome lattice, numbered $1,2,3$ in Table \ref{tab:latt}), so as to work with a $48 \times 48$ snapshot.

\subsection{Algorithms}

Here we give details on the two algorithms used in this work: Principal Component Analysis (PCA) and Convolutional Auto Encoder (CAE).  In the Appendix we discuss the effect of choosing other parameters and architectures for these algorithms.

\subsubsection{Principal Component Analysis}

Principal Component Analysis \cite{Bishop, Geron} is an algorithm usually used for dimensionality reduction. The idea is to find the hyperplane that is closest to the dataset, and project the data onto it. The different principal components are the succesive axes where the variance is preserved, and there are as many as the dimension of the dataset. Dimensionality reduction to $n$ dimensions is done projecting the first $n$ principal components. The number of such components may be chosen as to obtain a large enough accumulated variance, which is the sum of the dataset's variance that lies in each of the $n$ components. We chose to decompose the 5400 snapshot dataset in $n=500$ principal components, using a Gaussian kernel, where the $99\%$ of the variance is accumulated. In some studies, the two first principal components are used to explore and analyse the data. Here, we take the PCA with $d=500$ components that was trained with the skyrmion lattice dataset and use it to  reconstruct  snapshots obtained from simulations for  different models, temperatures and magnetic field. Then, we calculate the root mean square error between the input snapshot and the output one, as Eq.~(\ref{eq:rmse}), and explore whether this error may give us information on the physics of the models. Calculations were done using the SciKit Learn Python package \cite{SciKit}.

\begin{figure*}[th!]
\includegraphics[width=0.75\textwidth]{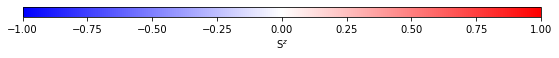} 
\includegraphics[width=0.75\textwidth]{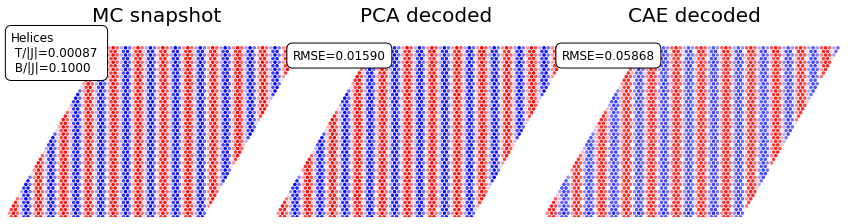} 
\includegraphics[width=0.75\textwidth]{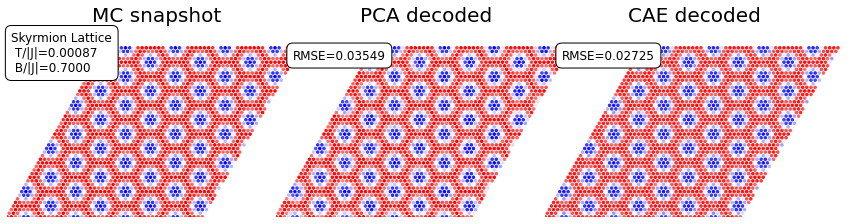} 
\includegraphics[width=0.75\textwidth]{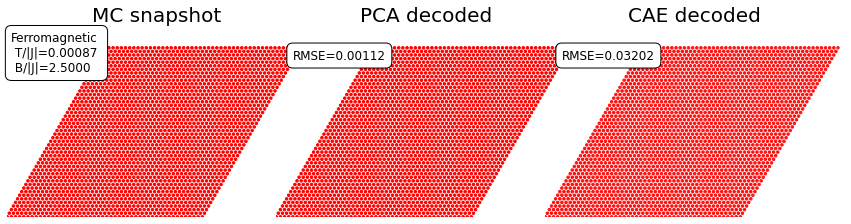} 
\includegraphics[width=0.75\textwidth]{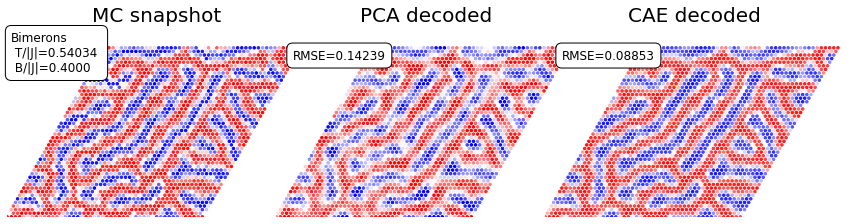} 
\includegraphics[width=0.75\textwidth]{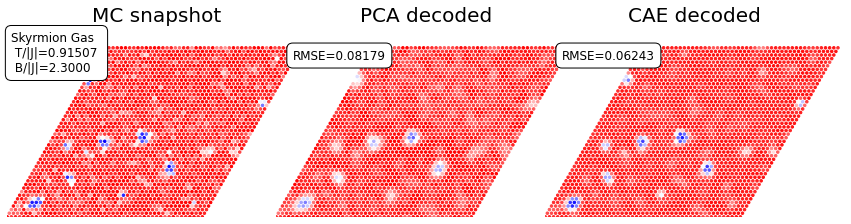}
\includegraphics[width=0.75\textwidth]{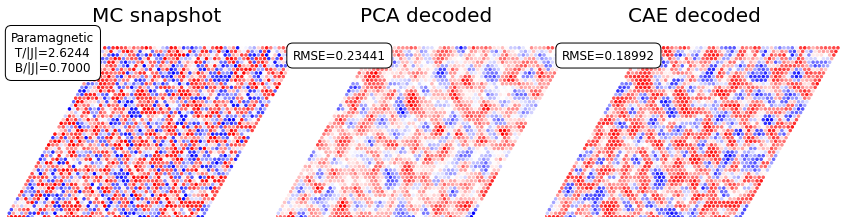} 
\caption{Examples of MC Snapshots and their corresponding decoded counterparts using the trained PCA and CAE for a simple ferromagnetic model in the triangular lattice (TFDM$^{xy}$,  in Table \ref{tab:latt}).  The first three rows show the three ordered phases that are found at very low temperature: helices, skyrmion lattice and ferromagnetic. Below, there are two types of phases that are enhanced by thermal fluctuations, the bimeron phase, found when helices get broken into elongated skyrmions or bimerons as the magnetic field is increased,  and the skyrmion gas, which emerges at higher fields, close to the ferromagnetic phase. The last row corresponds to a high temperature paramagnetic phase. }
\label{fig:snaps}
\end{figure*}

\subsubsection{Convolutional Autoencoder}

Autoencoders (AE) are algorithms where the data is compressed or encoded into a lower dimension space (called latent space) and then it is decompressed or decoded \cite{Geron}. If the decoded data retains the most relevant features of the input data, which may be measured with some quantity, then the AE has performed satisfactorily. Moreover, the compressed data in the latent space may be used for a certain application instead of the input data, with the advantage that it occupies less memory. Thus, one of the uses of AE is called ``data reduction''. However, AE has other uses; for example, since spurious features are removed in the process, the output data is ``cleaner'', and the AE may be used as a ``denoiser''. 

In our work, we will use the ``anomaly detection'' approach. The idea is to train an AE for a given type of data, and then apply it to a larger dataset, comparing the input and the output with a convenient variable, such as the root mean square error. If the error is small, then the input and output are similar, if it is large, then the input data may not be of the same type as that used in training the AE, so a larger error signals an anomaly. As in the PCA, we will use the artificial skyrmion lattice dataset to train a convolutional autoencoder (CAE -  an AE with a convolutional layer), measuring the RMSE. In this case, 80$\%$ of the data were used for training and 20$\%$ for validation. The RMSE for the trained CAE was 0.0178 in the training set and 0.0181 in the validation set.

The structure of the CAE is the following: the encoder consists of one convolutional layer with  activation function `relu',  32 filters, filter size 3 and padding `same', followed by a MaxPooling layer with filter size 2 and padding `same'. The decoder consists of a convolutional layer, with the same characteristics as the first one, and an UpSampling layer. It finalizes with a convolutional layer with one filter of size 3, a sigmoid activation function and padding `same'. The CAE was trained and validated with batch size 64, using early stopping for a maximum of 300 epochs with a patience of 10. Implementation of the CAE was done in TensorFlow using Keras \cite{Keras,TensorFlow}.

\section{Results and discussion}

In this section, we present the analysis of the comparison between the input and output data for the MC snapshots obtained when applying the PCA and CAE algorithms that were trained with the artificial set of  skyrmion lattices. We start in Subsec. A focusing on the three ferromagnetic models introduced in Sec. II (Table \ref{tab:latt}). Then, we will explore whether the anomaly detection technique using the same skyrmion-lattice trained CAE is able to provide some insight in two different antiferromagnetic models, which we will describe in Subsec. B.

\subsection{Application to ferromagnetic skyrmion models}

First, we compare the RMSE obtained for both algorithms for the TFDM$^{xy}$ snapshots. In order for the method to work, we would expect the RMSE to be lower at the skyrmion lattice phase, which is found at intermediate magnetic fields and lower temperature, and higher in the other phases, such as the low field helical phase at lower magnetic fields, or the high temperature paramagnetic phase. In Fig.~\ref{fig:snaps}, we present a series of examples comparing the input Monte Carlo snapshot with the output (decoded) snapshot after applying trained PCA and CAE. The first five rows correspond to  the different phases that arise  at lower and intermediate temperatures
as the magnetic field $B$ is increased:  helices, bimerons, skyrmion lattice, skyrmion gas and ferromagnetic.  At first glance, the decoded images seem very close to the input ones, both for PCA and CAE, even for the helical and the ferromagnetic cases, which are types of configurations that were not used to train the algorithms. In these examples, the RMSE (indicated in each decoded image) for PCA is even lower in the helical and ferromagnetic cases than in the skyrmion lattice. This is not ideal, since we are proposing to use a larger RMSE as an indicator that the system is further from an ordered skyrmion phase. However, there is a significant difference in the output images at higher temperatures, illustrated in the last row of Fig ~\ref{fig:snaps} for a snapshot in the paramagnetic phase: both the PCA and the CAE have a significantly higher RMSE.

\begin{figure}[th!]
\includegraphics[width=0.95\columnwidth]{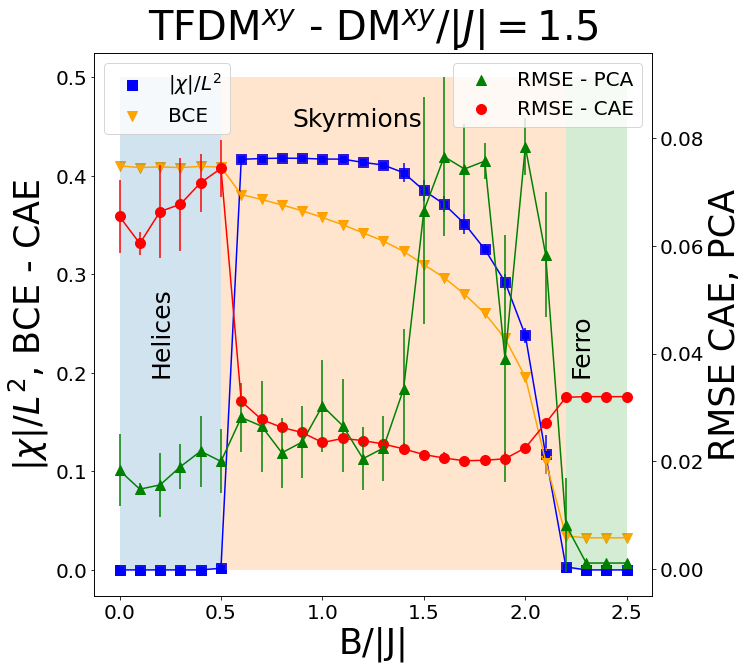}
\includegraphics[width=0.95\columnwidth]{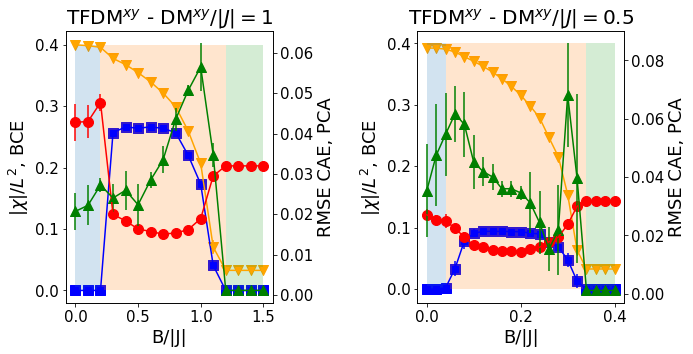} 

\caption{Different variables as a function of the external magnetic field $B$ at the lowest simulated temperature for simple ferromagnetic skyrmion model in the triangular lattice  (TFSM$^{xy}$)  for three different values of the in-plane DMI strength (top panel, $D^{xy}/|J|=1.5$, bottom panels $D^{xy}/|J|=1,0.5$): RMSE obtained comparing the MC input and the PCA and CAE outputs,  BCE calculated from the CAE outputs and the absolute value of the scalar chirality density$|\chi|/L^2$ obtained from the MC simulations.  The errorbars are the standard deviation of the quantities averaged over 10 independent copies. The background colors are a guide to the eye to indicate the three main phases of this model   (helices, skyrmions  and ferromagnetic).  }
\label{fig:TFcurves}
\end{figure}

\begin{figure}[th!]
\includegraphics[width=0.95\columnwidth]{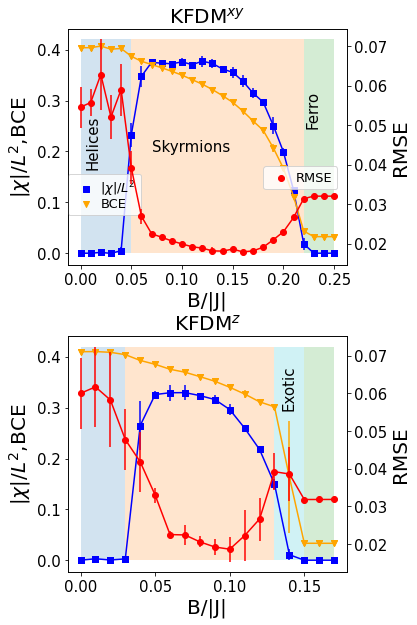} 

\caption{ RMSE and BCE from CAE, and scalar chirality density $|\chi|/L^2$  from MC simulations as a function of the external magnetic field $B$ at the lowest simulated temperature  ($T/|J|=0.0002$) averaged over 10 copies for two ferromagnetic skyrmion models in the kagome lattice: one with only in-plane DMI (KFDM$^{xy}$, top panel) and one with an additional out-of-plane DMI with the specific value  $D^z/|J|=\sqrt{3}$ (KFDM$^{z}$, bottom panel). The background colors are a guide to the eye to indicate the three main phases of the simple skyrmion models (helices, skyrmions and ferromagnetic), and the region in magnetic field where an exotic phase emerges in the KFDM$^{z}$ model, bottom panel.  }
\label{fig:2KFcurves}
\end{figure}

To further explore and quantify whether the RMSE obtained with PCA or CAE is a useful tool to distinguish between the low temperature phases, in Fig.~\ref{fig:TFcurves}, (top panel) we present the chirality from MC simulations,  PCA and CAE RMSE and the BCE as a function of the external magnetic field $B$ at the lowest simulated temperature for the TFDM$^{xy}$ model  ($T/|J|=0.0009$). The curves are averaged over 10 independent copies, and the error bars are calculated as the standard deviation.   Notice that this is the same model that was presented in Fig.~\ref{fig:skyPD}, and thus the chirality is the same as the one shown in the bottom panel of that figure. Let us recall that the chirality cannot be ``learned'' by the algorithms, since in this proposal only the $z$ spin component is used for training.

Turning to the  RMSE, comparing the RMSE calculated from the output images of both algorithms, it can be seen that the RMSE from PCA is in fact lowest for the ferromagnetic phase,  highest for the skyrmion gas, and, within the error, does not significantly distinguish between the helical and skyrmion lattice phases. On the other hand, the RMSE obtained comparing the MC snapshots and the CAE decoded ones shows a more interesting behaviour, clearly distinguishing three regions: a low field one with higher RMSE, corresponding to the helical phase, an intermediate one with the lowest values (the skyrmion phases), and then a third region where the RMSE goes slightly but noticeably up with $B$ and flattens at the ferromagnetic phase.   The BCE curve also seems to distinguish between these three regions, showing an abrupt change as  the system enters the skyrmion phases, but it is lowest in the ferromagnetic phase, and not in the skyrmion region, where the CAE was trained. Therefore,  given the behavior of the  RMSE for different magnetic fields, we choose the RMSE from the CAE to apply the anomaly detection technique in other models.  We emphasize that, regardless of the values of the obtained RMSEs and BCEs, in this work we are interested in using them to distinguish phases, and therefore we focus on the shape of the curves. As a comment, a closer inspection of the RMSE curve also suggests that the  RMSE error may also be a useful variable, an idea we will shortly come back to.

 To explore whether the overall behavior of these variables is independent of the $D^{xy}/J$ ratio (and thus, from skyrmion size), in the range where skyrmions are stabilized, in the bottom panels of Fig.~\ref{fig:TFcurves}, we present the same analysis  changing the values of DMI,  $D^{xy}/|J|=1,0.5$ in the TFDM$^{xy}$ Hamiltonian. A first observation is that the maximum value of the chirality density is smaller, simply because for smaller  $D^{xy}/|J|$ there are larger, and thus fewer, skyrmions for the same lattice size. We find that the behaviour of the variables obtained from CAE, RMSE and BCE, is qualitatively the same. For the RMSE from PCA, we see that it increases  at lower fields for $D^{xy}/|J|=0.5$,  since in this case, upon a small magnetic field the helices are easily distorted into bimeron type textures.  

\begin{figure}[th!]
\includegraphics[width=0.95\columnwidth]{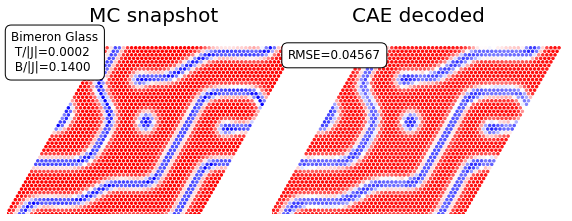} 
\caption{ MC Snapshots and their corresponding CAE decoded counterpart for the skyrmion model in the kagome lattice with an additional out-of-plane DMI  (KFDM$^z$) at a low temperature ``bimeron glass'' phase at high fields.  }
\label{fig:snapsKDz1}
\end{figure}

\begin{figure}[th!]
\includegraphics[width=0.95\columnwidth]{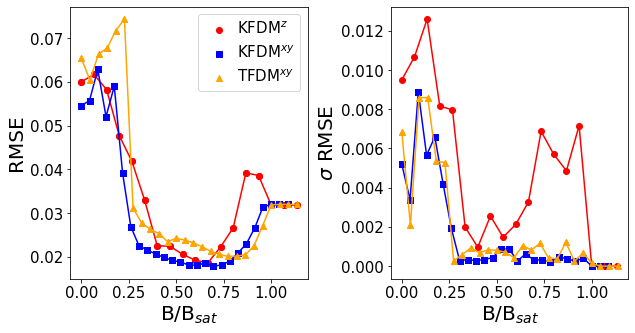} 

\caption{Mean value (left panel) and standard deviation (right panel) of the CAE RMSE as a function of magnetic field  for the three ferromagnetic skyrmion models introduced in Table \ref{tab:latt}  at the lowest simulated temperature. The magnetic field  was scaled to the saturation value for each model, $B_{sat}$, to facilitate the comparison}
\label{fig:RMS3F}
\end{figure}

Having chosen  our variable, the RMSE between the MC input snapshot and the decoded output from the trained CAE, we now apply the CAE to the other two ferromagnetic skyrmion models, both in the kagome lattice with in-plane interactions (KFDM$^{xy}$), and one with an additional out-of-plane DMI (KFDM$^{z}$) (see Table~\ref{tab:latt}). We plot the RMSE BCE and show the MC chirality density  as a function of magnetic field at the lowest simulated temperature for both models in Fig.~\ref{fig:2KFcurves}. Given that we are choosing one kagome sublattice to apply the CAE, which it is trained for triangular lattices,  in the rest of this work the chirality presented  for the kagome models is the sublattice chirality, calculated in the chosen triangular sublattice.  Notice that the chirality obtained from MC simulations is very similar for both models in the kagome lattice,  suggesting that the number of skyrmions and skyrmion size may be the same in both models, and gives no indication of exotic phases. However, this is not the case when exploring the RMSE curve. For the KFDM$^{xy}$, left panel, the behavior is very similar to the triangular lattice case. However, for the KFDM$^{z}$ model, before flattening in the ferromagnetic phase, the RMSE goes clearly up, there is a sharp jump in the BCE,  but,  as already mentioned, no significant differences are seen in the chirality, which is the order parameter connected to the skyrmion phases. Thus, the RMSE suggests that there is a different phase at higher magnetic fields, for   $B/|J|\sim 0.13-0.14$. Inspection of the snapshots reveals this is indeed the case; an example is shown in Fig.~\ref{fig:snapsKDz1}. It is an unusual high field bimeron + skyrmion phase, with possible glassy behaviour \cite{CSLSky2023}.  As mentioned in Sec.\ref{sec:model}, bimerons are usually found as intermediate textures between the helices and the formation of skyrmions, at lower magnetic fields. Regardless of the nature of the textures, the RMSE indicates that at higher fields the KFDM$^{z}$ presents a different phase than the other two ferromagnetic models. We compare the mean value and the standard deviation of the RMSE as a function of the external field (scaled to its saturation value in each model, $B_{sat}$)  for the three models in Fig.~\ref{fig:RMS3F}.  Inspecting the mean value of the RMSE presented in the left panel, it is clear that the RMSE is quantitavely and qualitatevely very similar in the three models in three regions (corresponding to helices, skyrmions and ferromagnetic), but evidently there is an exotic phase for the KFDM$^{z}$ case before the system goes into the ferromagnetic phase. The jump of the RMSE for this model in this region is not seen in the other two cases, strongly supporting the idea that in that region of $B/B_{sat}$ the system goes into a different type of ordering.  Interestingly, this information may also be extracted from the standard deviation of the RMSE ($\sigma$), seen in the right panel of Fig.~\ref{fig:RMS3F}: in the  TFDM$^{xy}$ and  KFDM$^{xy}$ models, $\sigma$ is significantly larger at small magnetic fields, goes down at intermediate fields and flattens to zero at higher fields. For  KFDM$^{z}$, $\sigma$ goes up again before going to zero, signaling a different behavior in this model.

\begin{figure}[th!]
\includegraphics[width=0.95\columnwidth]{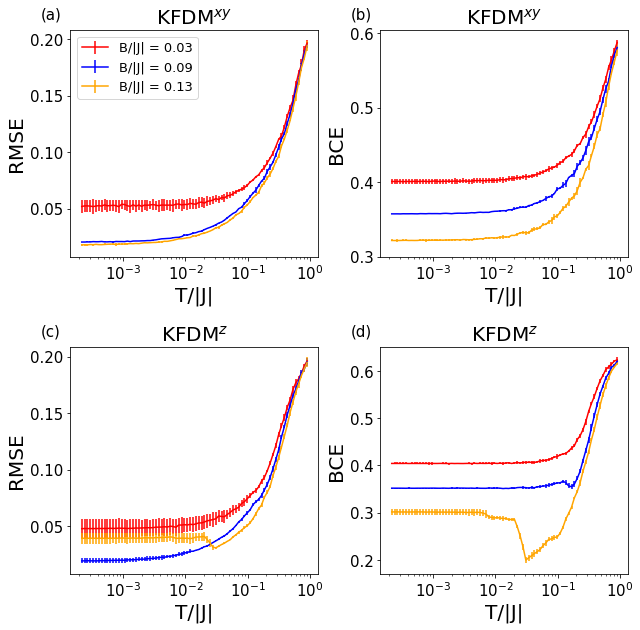} 

\caption{RMSE and BCE as a function of temperature for three magnetic fields  $B/|J|=0.03,0.09,0.13$ for the two skyrmion ferromagnetic kagome models, with in-plane DMI, KFDM$^{xy}$ (panels (a,b)) and with in-plane and an additional out-of-plane DMI, KFDM$^{z}$ (panels (c,d))  }
\label{fig:2KFvsT}
\end{figure}

\begin{figure}[th!]
\includegraphics[width=0.95\columnwidth]{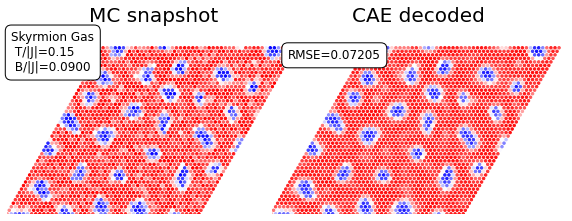} 
\caption{ MC Snapshots and their corresponding CAE decoded counterpart for the skyrmion model in the kagome lattice with an additional out-of-plane DMI  (KFDM$^z$) at  an intermediate temperature skyrmion gas in a field polarized background with chiral spin liquid properties.  }
\label{fig:snapsKDz2}
\end{figure}

\begin{figure*}[th!]
\includegraphics[width=0.95\textwidth]{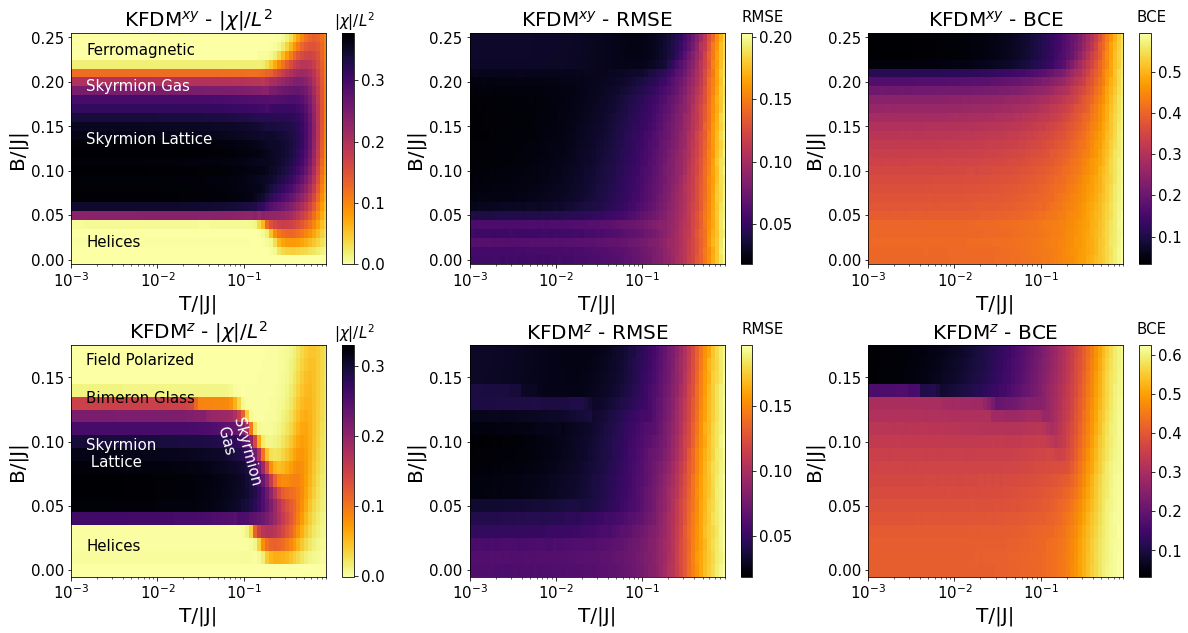} 

\caption{  Phase diagrams for as a function of temperature and magnetic field for the two skyrmion ferromagnetic kagome models, with in-plane DMI, KFDM$^{xy}$ (top row) and with in-plane and an additional out-of-plane DMI, KFDM$^{z}$ (bottom row) of the chirality density obtained from Monte Carlo simulations (left column), and the RMSE and BCE calculated between the input and output snapshots of the CAE (middle and right columns). }
\label{fig:pd}
\end{figure*}

The KFDM$^{z}$ model has been shown to have several exotic behaviors, due to the competition between skyrmion and chiral spin liquid physics \cite{CSLSky2023,CSLSky2024}. Besides the ``bimeron glass'' above mentioned, for the value of $D^{xy}$ chosen here, it has been shown that, coming from the high temperature paramagnetic phase, the skyrmion lattice is formed by the discrete emergence of skyrmions at intermediate temperatures, forming a skyrmion gas  with well formed skyrmions in a field polarized background with chiral spin liquid properties  that gets more densely populated as the temperature is lowered. To explore whether this analysis can shed light on this phase, in Fig. ~\ref{fig:2KFvsT} we compare the RMSE and BCE as a function of temperature  for the two ferromagnetic kagome skyrmion models, at three different values of the magnetic field  $B/|J|=0.03, 0.09, 0.13$. In panels (a) and (b) it can be seen that for the KFDM$^{xy}$ model both RMSE and BCE    go smoothly down to a given value as the temperature is lowered. This given value for BCE gets smaller as the magnetic field is increased (panel (b)), but for the RMSE (panel (a)) the low temperature value   is similar for both  $B/|J|=0.09, 0.13$. This is simply due to the fact that for this model at those magnetic fields the system is in a skyrmion crystal phase (see Fig.~\ref{fig:2KFcurves}, left panel). The variables hint a different physical process   with temperature for the KFDM$^{z}$ model, Fig. ~\ref{fig:2KFvsT}, panels (c) and (d). Both the RMSE and BCE curves show a sharp feature at  $B/|J|=0.13$, coming from a higher value at higher temperatures  ($T/|J|\sim1$), they go down with temperature, show a minimum at intermediate temperatures  ($T/|J|\sim0.002$) and then jump to settle to their low temperature value. A similar feature is seen for the BCE at  $B/|J|=0.09$. 

As discussed in previous works \cite{CSLSky2023,CSLSky2024}, for these $B$ the KFDM$^{z}$ model goes through an  skyrmion gas phase at intermediate temperatures, with a field polarized chiral spin liquid background. Clearly this type of phase is better decoded by the CAE than the bimeron glass phase found at  lower temperatures,    and therefore both the BCE and RMSE show a minimun at intermediate temperature. We present an example MC snapshot of this phase and its decoded image in Fig.~\ref{fig:snapsKDz2}. Since the temperature is higher and the background is a  field polarized chiral spin liquid, it can be seen that it is not completely aligned with the field (i. e. the background is not completely red). On the other hand, this changes when decoding the image with the CAE, where thermal fluctuations are smoothed. This is related to the potential ``denoising'' use of convolutional autoencoders, showing that these algorithms, if needed, may also be considered as tools to ``erase'' fluctuations and better define structures such as skyrmions.

As a summary, we present a comparison of the $B-T$ phase diagrams of the  chirality density obtained from MC simulations, and the calculated RMSE and BCE from CAE for the two kagome models, in Fig.~\ref{fig:pd}. For both models, we first see that the RMSE and BCE are higher at higher temperature, as expected. Then,  for BCE, at the lower temperatures it goes down with magnetic field and is lowest when the system is in the field polarized phase, and it does not seem to easily distinguish the lower field phases. However, at intermediate temperatures we see a particular behaviour in the KFDM$^{z}$ case: a drop in the BCE, which matches the chiral spin liquid region \cite{CSLSky2023,CSLSky2024}. As for the RMSE, in the KFDM$^{xy}$ model at lower temperatures it distinguishes between three regions, with a higher value at lower magnetic field (corresponding to helices and bimerons), lowest values at intermediate magnetic fields (skyrmions phase) and then goes a bit up at higher fields, in the field polarized region. An important difference is observed for KFDM$^{z}$: a small region at higher field where the RMSE goes up again, between the skyrmion lattice phase and the field polarized one. This is another exotic feature of this model, a higher field bimeron phase with potentially glassy characteristics \cite{CSLSky2023}, which is not evident inspecting the chirality. Thus, we see that the RMSE, and also BCE, are powerful tools to pinpoint regions in parameter space where a system may deviate from the ``typical'' skyrmion phase diagram. Given that we only analyse snapshots, this technique may be applied to real-space images, and provide helpful insight of a skyrmion system without resorting to designing a model and permorfing simulations.

\subsection{Application to antiferromagnetic models}

Although the CAE was trained on ferromagnetic  skyrmion lattices, in this subsection we aim to explore whether this CAE may give some insight in two well-known antiferromagnetic models in the triangular lattice, presented in Table \ref{tab:lattAF}. For these models, the exchange interaction is antiferromagnetic ($J>0$), which implies a frustrated system due to the lattice geometry. In the first model, TAFDM$^{xy}$, there are in-plane DMI ($D=D^{xy}\neq 0$). It has been shown that under a magnetic field, at finite temperature an antiferromagnetic skyrmion lattice formed by three interpenetrated sublattices is stabilized \cite{Rosales2015}, a texture that has been shown to be stabilized in other frustrated models  \cite{Mohylna2022,Mohylna2021,Mohylna2022JMMM}, and connected to the fractional antiferromagnetic skyrmions in MnSc$_2$S$_4$ \cite{Gao2020,Rosales2022}. As a second model we consider the pure exchange antiferromagnetic model  ($J>0,D=0$), where the Hamiltonian may be rewritten as a sum of the total spin of triangular plaquettes.  Under a magnetic field $B$ , a notable feature of this model is that at  $B/|J|\sim 3$ at finite temperature order-by-disorder (entropic selection) induces a pseudo-plateau at magnetization $M=1/3$, where in each plaquette two spins are parallel to the external field (``up'', $u$) and one is antiparallel (``down'', $d$), and is thus known as a $uud$ pseudo-plateau \cite{Gvozdikova2011,Seabra2011,Kawamura1985}. For lower fields, there is a coplanar ``Y'' state, and for higher fields a coplanar ``V'' state \cite{Gvozdikova2011}.

\begin{table}[th!]
\centering
\begin{tabular}{|c|c|c|}
\hline 
& TAFDM$^{xy}$ & TAF  \\
\hline 
\raisebox{-.1\normalbaselineskip}[0pt][0pt]{\rotatebox[origin=c]{90}{model}} & \begin{minipage}{.2\textwidth}
      \includegraphics[width=0.57\linewidth]{fig_tab1_a.png}
    \end{minipage} & \begin{minipage}{.2\textwidth}
      \includegraphics[width=0.57\linewidth]{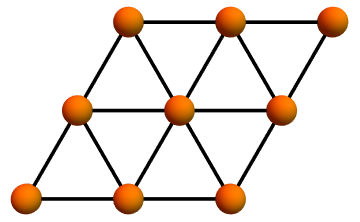}
    \end{minipage} \\
     
\hline 
 $J/|J|$ & 1 &  1 \\
\hline
$D^{xy}/|J|$ & 0.5 &  0  \\
\hline 
\end{tabular}
\caption{\label{tab:lattAF} Two well known antiferromagnetic (AF) models in the triangular lattice. Both have antiferromagnetic exchange interactions  $J/|J|=1$, one has in-plane DMI (TAFDM$^{xy}$) and the other one is the pure exchage model (TAF) }
\end{table}

Clearly, the  emergent antiferromagnetic low-temperature textures from these models are quite different from the ferromagnetic skyrmion lattices, and thus a significantly larger RMSE is expected. To explore whether some information on these models may nonetheless be extracted with the skyrmion-trained CAE, we proceed as before: we apply the CAE to low temperature snapshots obtained from MC simulations of both models, and calculate the RMSE and the BCE. 

\begin{figure}[th!]
\includegraphics[width=0.95\columnwidth]{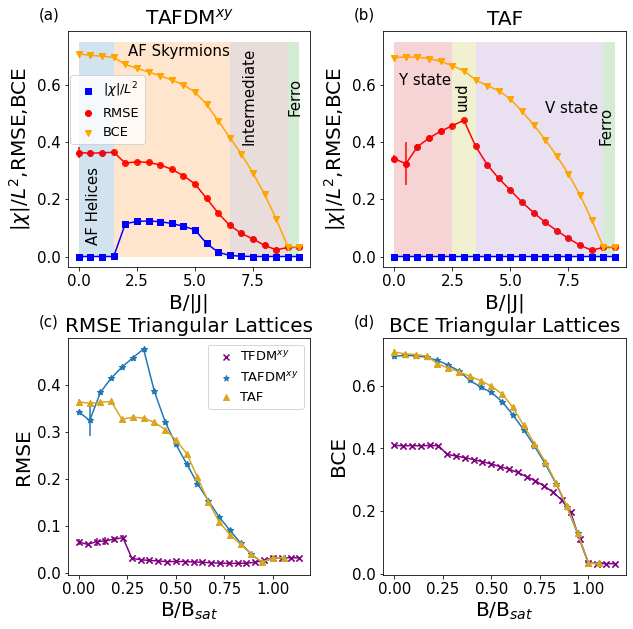} 

\caption{RMSE and BCE calculated after applying the CAE and chirality from MC simulations as a funcion of the external magnetic field $B$ at the lowest simulated temperature $T/|J|=0.0009$) for  TAFDM$^{xy}$ (panel (a))  and the TAF (panel (b)) models. In both panels, the shaded background colors are guides to the eye to indicate the different phases. Comparison of the CAE (panel (c)) and BCE (panel (d)) as a function of $B$ for the three models in the triangular lattice: the ferromagnetic skyrmion model TFDM$^{xy}$, the antiferromagnetic skyrmion model TAFDM$^{xy}$  and the pure exchange antiferromagnetic case TAF.  In panels (c) and (d) the magnetic field  was scaled to the saturation value for each model, $B_{sat}$, to facilitate the comparison.}
\label{fig:2TAF}
\end{figure}

In Fig.~\ref{fig:2TAF} (a) and (b) we show the RMSE, BCE and chirality $\chi$ as a function of the external field $B$ for the TAFDM$^{xy}$ and TAF models, respectively, at the lowest simulated temperature  $T/|J|=0.001$). In the TAFDM$^{xy}$ case (panel (a)),  the chirality curve is similar to that from the TFDM$^{xy}$ models in Fig.~\ref{fig:TFcurves}, and therefore one might expect similar low temperature magnetic textures from this MC order parameter. Inspecting the RMSE and BCE curves, there is a strong hint that something different is going on, since the errors are higher and the curves as a function of  $B/|J|$ are not the same. The RMSE and the BCE  are relatively flat at lower  $B/|J|$ and  present a sharp decrease  at  $B/|J| \sim 2$ , which matches the jump in the chirality $\chi$ associated with the stabilization of the AF skyrmion lattice. Then as  $B/|J|$ is increased both quantities go down, and the curves get steeper when $\chi$ drops to zero. Although the behaviour of the RMSE and the BCE is similar, the features in the RMSE are sharper than for BCE, particularly the steep drop at lower field. So, in this case, the RMSE shows that there is a first type of phase stabilized at  $B/|J|>2$, followed by a different type of phase up to  $B/|J| \sim 4$, where the RMSE starts to drop until it flattens at its minimum value at higher fields, associated to the ferromagnetic phase, with all spins aligned with $B$. Moreover, since it is not at a minimum where the chirality is highest, it also suggests that the textures in these range of magnetic field are not ferromagnetic skyrmion lattices, something that is not obvious by simply inspecting the chirality curve from MC.

\begin{figure}[th!]
\includegraphics[width=0.95\columnwidth]{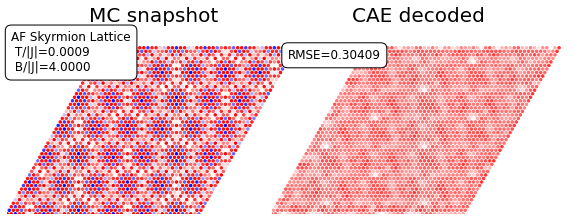} 
\includegraphics[width=0.95\columnwidth]{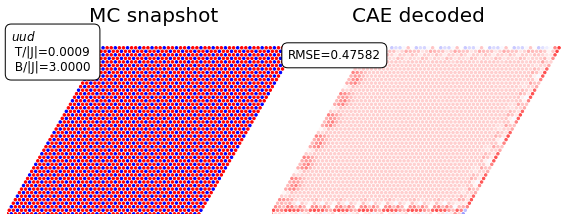} 
\caption{  Example of MC snapshots and their corresponding CAE decoded counterparts  of the antiferromagnetic skyrmion phase  in the antiferromagnetic triangular lattice at low temperature (TAFDM$^{xy}$, top row) and  and of the $uud$ $M=1/3$ pseudo-plateau in  the pure exchange antiferromagnetic model in the triangular lattice (TAF, bottom row).}
\label{fig:snapsTAFs}
\end{figure}

 The BCE and RMSE curves are quite different for the TAF case (panel (b)): the BCE  goes down with increasing magnetic field, presenting a small kink at  $B/|J|\sim 3$, the  pseudo-plateau magnetic field. On the other hand, the RMSE curve has two clearly distinct regions, separated at  $B/|J|\sim 3$: a low magnetic field region where the RMSE goes up, and a higher field region where it goes down, presenting thus a peak at  $B/|J|\sim 3$. Therefore, even if, as expected since the CAE was trained for skyrmion lattices, the obtained RMSE values are quite large, the RMSE calculated after applying this CAE to the MC snapshots indicates that there are two types of phases at lower and higher field, and that there is a particular feature in the TAF model at  $B/|J|\sim 3$,  since the error here is highest and the curve is sharp.

Panels (c) and (d) show  the RMSE and BCE curves, respectively, for these two AF models and the simple ferromagnetic skyrmion model TFDM$^{xy}$; here the magnetic field $B$ has been (scaled to its saturation value in each model, $B_{sat}$)   for a better comparison of the magnitude and characteristics of the curve . As expected, the magnitude of both quantities, particularly RMSE, is significantly larger for the antiferromagnetic models. The three RMSE curves, contrary to the BCE ones, have different characteristics, showing that the three models  have different types of phases. The kinks and jumps in the three curves are also indicators of the different phases within each model.

Finally, in Fig.~\ref{fig:snapsTAFs} we show the input and output of a low temperature snapshot in each AF model. The first row corresponds to a three-sublattice antiferromagnetic skyrmion crystal (TAFDM$^{xy}$) and the second one to the pseudo-plateau at $M=1/3$ in the pure exchange model (TAF). Clearly, in both cases the decoded image is quite different from the input one. This is due to the fact that the CAE was trained with perfect ferromagnetic skyrmion sizes, where the textures are smoothly modulated, and here it is applied to two snapshots where there is an abrupt change in the texture at nearest neighbor distances, given the antiferromagnetic nature of the models. The RMSE is largest at the $uud$ pseudo-plateau, which is  completly washed away by the CAE, trained for more ``coarse-grained'' textures.

 \subsection{Application to the square lattice}

\begin{figure}[h!]
    \centering
    \includegraphics[width=0.9\columnwidth]{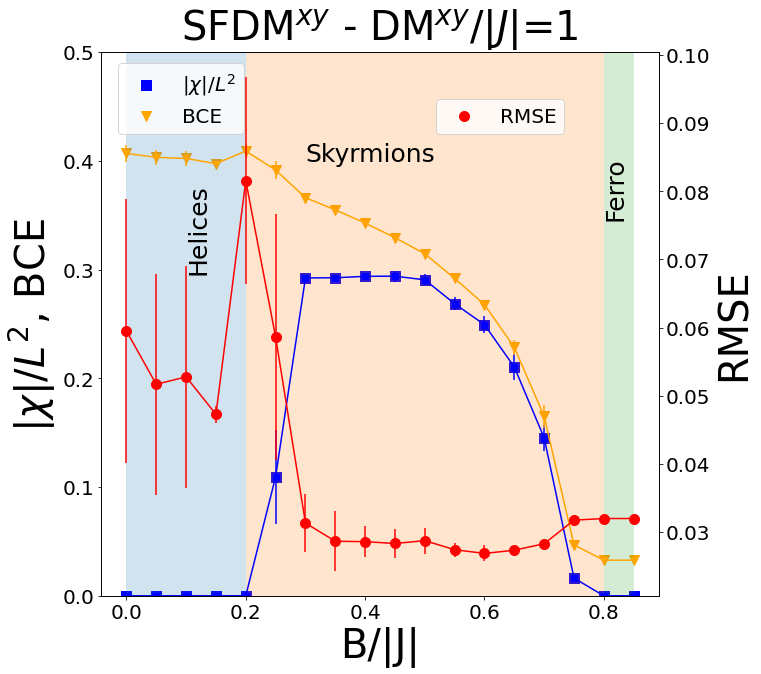}
    \caption{ RMSE and BCE from CAE, and scalar chirality density $|\chi|/L^2$  from MC simulations as a function of the external magnetic field $B$ at the lowest simulated temperature ($T/|J|=0.00087$) averaged over 5 copies for a ferromagnetic skyrmion model in the square lattice with in-plane DMI, $D^{xy}/|J|=1$.  The background colors are a guide to the eye to indicate the three main phases of the simple skyrmion models, as presented in Figs.~\ref{fig:TFcurves} and \ref{fig:2KFcurves}} 
    \label{fig:SFcurves}
\end{figure}

Throughout  this work, we have focused on triangular based lattices. We now explore the portability of the trained CAE, applying it to snapshots from simulations for the Hamiltonian in Eq.(\ref{eq:H}) in the $L=48$ square lattice for in-plane DMI  $D^{xy}/|J|=1$ (SFDM$^{xy}$). As in the previous cases, in Fig.~\ref{fig:SFcurves} we plot, for the lowest simulated temperature ($T/|J|=0.00087$),  the RMSE and BCE between the input and output images as a function of the external magnetic field, and we also show the chirality density calculated from simulations to analyze what information may be extracted from the behavior of the RMSE and BCE.

\begin{figure}[h!]
    \centering
    \includegraphics[width=0.8\columnwidth]{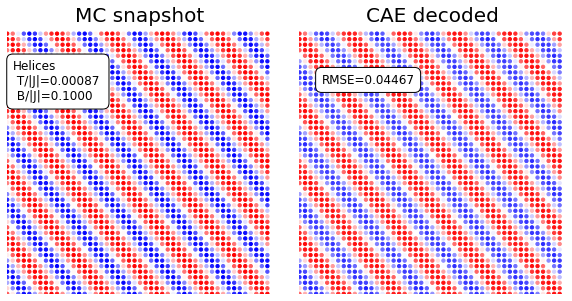}     \includegraphics[width=0.8\columnwidth]{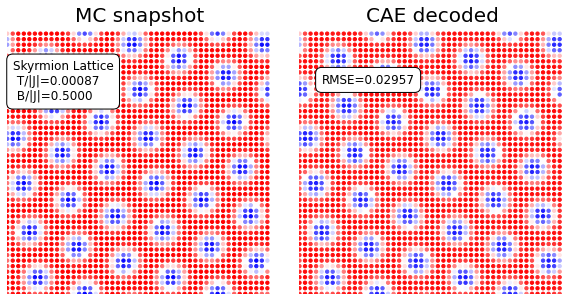}
    \caption{ Example of MC Snapshots and their corresponding CAE decoded counterparts for the skyrmion model in
the ferromagnetic square lattice SFDM$^{xy}$ at low temperature for the helical  (top row) and the skyrmion lattice phases (bottom row)} 
    \label{fig:sqsnaps}
\end{figure}

As expected, the behavior of the chirality in this lattice is as in the triangular lattice case: as a function of the magnetic field, there is an intermediate region of non-zero chirality, related to the formation of skyrmions. Surprisingly, we see that, even if we applied a CAE trained with triangular based lattices, both the BCE and the RMSE  follow a trend similar to that seen in the  TFDM$^{xy}$ models in Fig.~\ref{fig:TFcurves}. Most importantly, the RMSE is also minimum for the skyrmion phase, with a value comparable to the triangular lattice models. This may not be expected, since the CAE reads the input data as a matrix, and thus a skyrmions  in a triangular lattice are actually seen more elongated. In Fig.~\ref{fig:sqsnaps} we show two examples of input and output configurations for the square lattice, corresponding to a helical and a skyrmion lattice configuration, showing that the outputs retain the main features of these phases even in a different lattice geometry.

%%%%%%%%%%%%%%%%%%%%%
\section{Conclusions} 
%%%%%%%%%%%%%%%%%%%%%

In this work we propose the use of the anomaly detection technique to explore skyrmion phase diagrams. We trained two types of algorithms, Principal Component Analysis and Convolutional Autoencoder, with a data set of analytically generated skyrmion lattices. Our main idea was to compare  input configurations, obtained with Monte Carlo simulations from a given model, with the decoded (output) snapshot generated when applying these algorithms to the input data. A large deviation from the original MC data would imply that the chosen configuration is ``further'' from a skyrmion lattice. To quantify the difference between the input and output snapshots, we calculated the Root Mean Square Error for both PCA and CAE, and the Binary Cross Entropy for CAE. 

First, we chose a simple and well known skyrmion model in the triangular lattice, one that combines ferromagnetic exchange  and in-plane Dzyaloshinskii-Moriya interactions under an external magnetic field. At low temperature and low magnetic field, helices are stabilised. As the field increases, a skyrmion lattice is formed, which then gives place to a skyrmion gas until the spins are completely polarized. Comparing the RMSE from PCA and CAE, we found that for PCA the lowest RMSE was found for the ferromagnetic phase, and that, within the errorbars, it was not possible to distinguish between the helical and the skyrmion phases. On the other hand, the CAE RMSE showed a clear distinction between these three main phases, and was lowest in the skyrmion region, which is essential for out proposal. Surprisingly, the RMSE is not very large for helical and ferromagnetic phases, and the decoded snapshots are quite close to the input ones. We also compared the behaviour of the RMSE and the BCE. Although the BCE may be able to separate between the low temperature regions of this model, it monotonically goes down with the magnetic field, and thus is minimum for the ferromagnetic snapshots.

Secondly, we applied the CAE to two models in the kagome lattice, the well known skyrmion model we had simulated in the triangular lattice, and one where an additional out-of-plane DMI is included.  In both cases, the chirality from MC simulations, which is the order parameter for skyrmions, is quite similar. Regarding the errors, as expected, the behaviour of the RMSE in the first case (only planar DMI) is similar to that in the triangular lattice. However, in the second case (with both planar and out-of-plane DMI) there is a clear region at low temperature and higher fields where the error raises, indicating the possibility of a different type of phase(an ``anomaly''). In fact, previous studies in this model \cite{CSLSky2023} show that there is a high field bimeron glass phase. Moreover, this model also has a variety of phases at higher temperature, which is reflected in both the RMSE and BCE. Our analysis also suggests that the spread of the RMSE, considering independent realizations for the MC data, may also be a tool to distinguish between different low temperature phases. 

 Moreover, we apply the skyrmion-lattice trained CAE to snapshots from two models in the antiferromagnetic triangular lattice: one with in-plane DMI, where antiferromagnetic skyrmion lattices are stabilized, and the pure exchange model, which presents a well known $uud$ plateau at $M=1/3$. Given that the CAE is trained with ferromagnetic skyrmion lattices, a good decodification of the input data is not expected. Nonetheless, the low-temperature RMSE curves have features that suggest possible different phases. For the antiferromagnetic skyrmion model, the chirality curve does not indicate significant differences from the ferromagnetic case. However, the RMSE curves do, and the values of the RMSE are higher than in the ferromagnetic case, signaling that the resulting textures with non zero-chirality are not ferromagnetic skyrmion lattices. Then, in the pure antiferromagnetic exchange model, the RMSE is maximum for values of the magnetic field  where the pseudo-plateau emerges. 

 Finally, we explore the possibility of applying the  CAE trained with triangular lattices in other geometries. To do this, we repeated the analysis for a ferromagnetic skyrmion model in the square lattice, and showed that the RMSE from CAE is able to distinguish between the three main low temperature phases, thus expanding the possibilities of application.

In conclusion, we have shown how the anomaly detection technique, which we have applied resorting to a very simple Convolutional Autoencoder trained with analytically generated skyrmion lattices and choosing the RMSE to measure the error between input and output data, may give relevant information on skyrmion systems,  which may not be easily determined simply by studying the chirality parameter. We have demonstrated that the comparison and study of the RMSE was able to distinguish the three typical phases in a skyrmion phase diagram (helices, skyrmions and field-polarized spins), and to spot regions in parameter space where there may be exotic behavior. Additionally, we have seen how it may also hint the existence of different phases in other types of models, and that it may be applied to skyrmion models in other lattice geometries. Thus, we expect our work to support and further promote the exploration and use of machine learning techniques in different magnetic models.  On the one hand, these techniques complement the analysis from simulations, and, on the other, since snapshots are treated as images, these techniques may even be applied to experimental data  obtained with spin-polarized scanning tunneling microscopy techniques \cite{Yu2010,Bergmann2014,Hirschberger2019,Yasui2020}.

%%%%%%%%%%%%%%%%%%%%%
\section*{Acknowledgments} 
%%%%%%%%%%%%%%%%%%%%%

This work was partially supported by CONICET (PIP 2021-11220200101480CO, PIBAA 28720210100698CO), Agencia I+D+i PICT
2020 Serie A 0320 and SECyT UNLP PI+D X893 and X947.

%%%%%%%%%%%%%%%%%%%%%
\section*{Appendix}
%%%%%%%%%%%%%%%%%%%%%%%%%
\begin{figure}[b!]
    \centering
    \includegraphics[width=0.9\columnwidth]{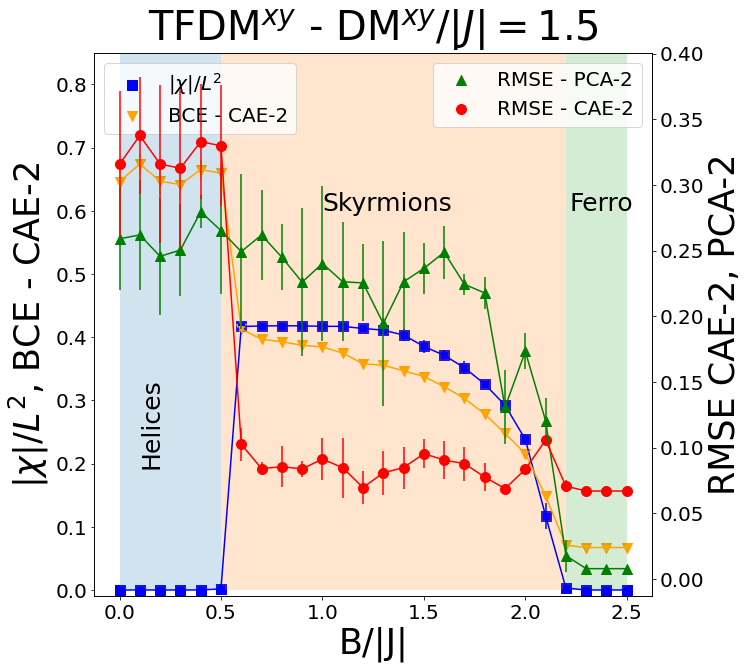}
    \caption{ Chirality density from MC simulations, calculated RMSE from a different PCA model (PCA-2), RMSE and BCE from a second CAE model (CAE-2) as a function of magnetic field at the lowest simulated temperature for the ferromagnetic skyrmion model in the triangular lattice (TFDM$^{xy}$).}
    \label{fig:PCA2}
\end{figure}

Here we discuss the effect of changing parameters of the PCA and CAE in our analysis. First, we focus on the PCA. Aiming for more variability, we reduce the number of principal components of the PCA from $n=500$ to $n=120$ in order to reach an  accumulated variance of $~80\%$  (PCA-2). Even if this may imply a larger RMSE, it could also imply larger differences between the helices and the skyrmion phases, for example, since it would retain the main features of skyrmion lattices. However, we find that this is not the case, we present the PCA-2 RMSE as a function of $B$ at the lowest simulated temperature for TFDM$^{xy}$ in Fig.~\ref{fig:PCA2}, to be contrasted with the upper panel of Fig.~\ref{fig:TFcurves}. Although the RMSE seems to be a bit larger in the helical phase, the differences between the helical and the skyrmion region are be enough to distinguish between the two phases within the errors.

\begin{figure*}[th!]
\includegraphics[width=0.95\textwidth]{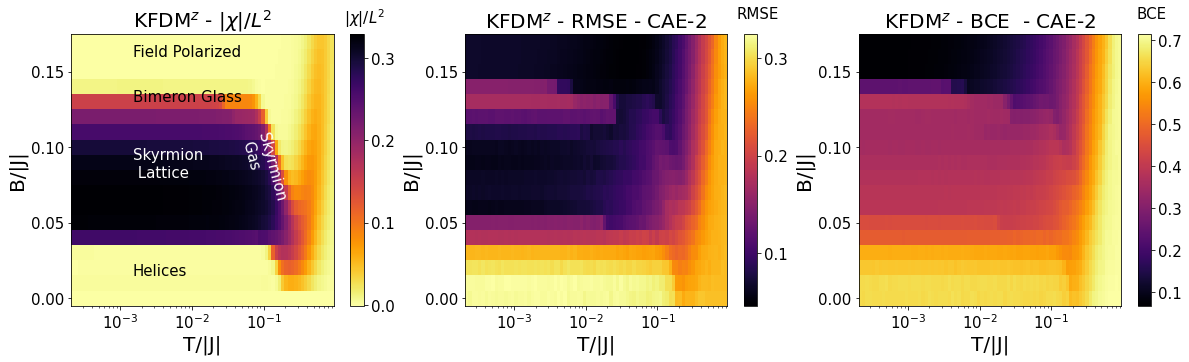} 
\caption{  Phase diagrams  as a function of temperature and magnetic field for the skyrmion ferromagnetic kagome model with in-plane and an additional out-of-plane DMI, KFDM$^{z}$  of the chirality density obtained from Monte Carlo simulations (left panel), and the RMSE and BCE calculated between the input and output snapshots of the second CAE model, CAE-2 (middle and right panels). }
\label{fig:pdcae2}
\end{figure*}

Secondly, we construct a different CAE. The CAE used in the main work is very simple and only has one convolutional and one MaxPooling layer for the enconding part. We construct a deeper CAE (CAE-2) adding in the encoder an extra convolutional layer with 64 filters of size 4 and  MaxPooling layer with filter size 3, and thus in the decoder we add a convolutional  and an UpSampling layer. The rest of the structure and the training method remains unchanged. We get a RMSE of 0.0425 in the training set, and 0.0432 in the validation set. To explore whether a change in the CAE affects the main result of this work, i. e. the possibility to distinguish phases in a skyrmion phase diagram and to detect regions with possible exotic textures  that may not be easily distinguished with MC order parameters, we apply the CAE-2 to the ferromagnetic models. Results for TFDM$^{xy}$ are shown in Fig.~\ref{fig:PCA2}: the RMSE is clearly different for the helices and the skyrmion phases, but the distinction is not that clear between the skyrmion phases and the ferromagnetic order. There is a single point with a slightly larger value, that corresponds to the skyrmion gas phase, which is narrow at lower temperatures.   The BCE retains its overall behavior (it is highest in the helices and smallest in the ferromagnetic case), although the jumps between the three main phases are sharper. 

\begin{figure*}[th!]
\includegraphics[width=0.95\columnwidth]{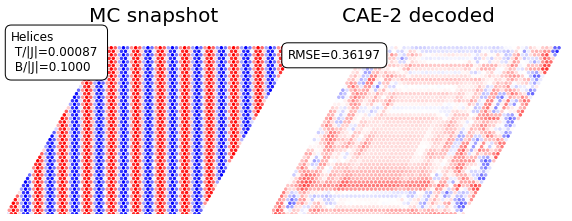}
\includegraphics[width=0.95\columnwidth]{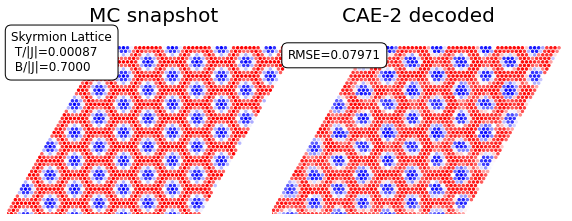}
\includegraphics[width=0.95\columnwidth]{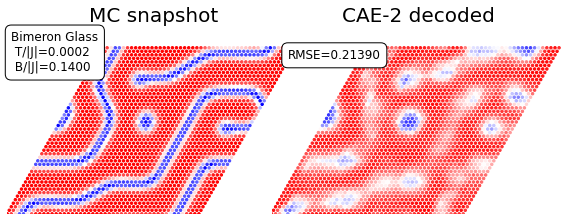} 
\includegraphics[width=0.95\columnwidth]{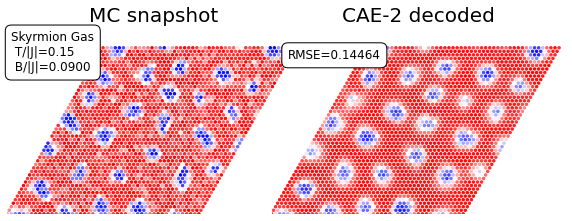} 
\caption{ MC Snapshots and their corresponding decoded counterpart using a deeper CAE model (CAE-2) for two typical phases of the skyrmion model in the triangular lattice (TFDM$^{xy}$, first two rows), and for two exotic higher field phases found in the kagome lattice with an additional out-of-plane DMI  (KFDM$^z$, last two rows). }
\label{fig:snapscae2}
\end{figure*}

To explore whether CAE-2 may detect exotic phases, we compare the results of the RMSE and BCE phase diagrams for the kagome model KFDM$^{z}$  in Fig.~\ref{fig:pdcae2}, to contrast with Fig.~\ref{fig:pd}.  We find that, in general, the RMSE is bigger, implying that the reconstruction is not as good as in the first proposed CAE. However,  different regions may more easily be distinguished in temperature and magnetic field, and the anomaly detection technique comparing the RMSE in different regions may still work.
Most interestingly,  the exotic phases (the high field bimeron glass and the intermediate skyrmion gas phases) correspond to larger changes both in the RMSE and in the BCE phase diagrams. However, the larger RMSE values are related to poorer reconstructions of the spin configurations, as we show below. In Fig.~\ref{fig:snapscae2} we present the MC snapshot and decoded images for helices and skyrmion lattice for the TFDM$^{xy}$ model (to be compared with Fig.~\ref{fig:snaps}) and for the bimeron glass and the skyrmion gas that emerge in the KFDM$^z$ case (to be compared with Figs.~\ref{fig:snapsKDz1},\ref{fig:snapsKDz2}). Clearly, the  CAE-2 fails to reconstruct features that are not skyrmions, and even in the skyrmion lattice case the decoded configuration looses definition when compared to the first CAE model,  giving the impression of  ``blurred'' images. Although a good reconstruction is not the key of the technique, the lack of significant differences in the RMSE in the skyrmion and ferromagnetic regions could imply that these types of phases may not be distinguished in other contexts. .

\begin{figure}[h!]
    \centering
    \includegraphics[width=0.9\columnwidth]{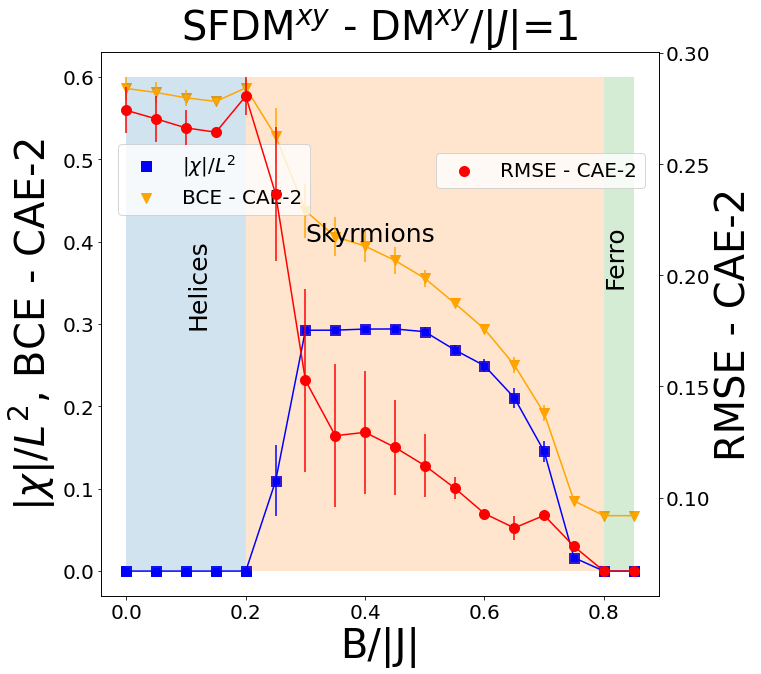}
    \includegraphics[width=0.45\columnwidth]{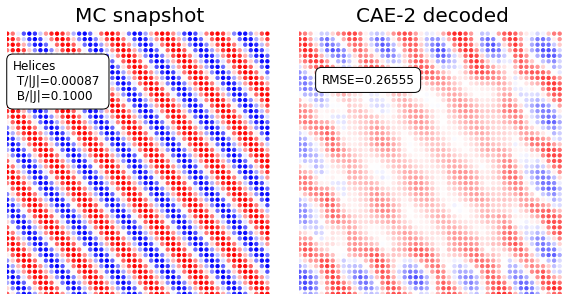}
    \includegraphics[width=0.45\columnwidth]{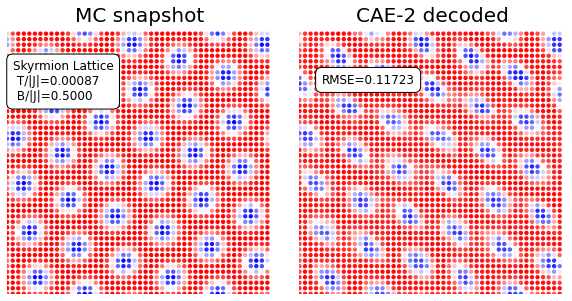}    
    \caption{(Top) Scalar chirality density from MC simulations, BCE and RMSE from a second CAE model as a function of magnetic field at the lowest simulated temperature and (bottom) MC snapshots and their CAE-2 decoded counterparts for the SFDM$^{xy}$ model}
    \label{fig:sqcae2}
\end{figure}

Finally, we apply the CAE-2 to the ferromagnetic square lattice model. Contrary to the ferromagnetic cases in triangular based lattices discussed above, we find that, although the BCE curve is similar to that of the TFDM$^{xy}$ model presented in Fig.~\ref{fig:PCA2},  here the RMSE  in the skyrmion phase is higher than in the ferromagnetic phase, as shown in the top panel of Fig.~\ref{fig:sqcae2} where we plot the BCE and the RMSE from CAE-2 as a function of magnetic field at the lowest simulated temperature, to be compared with Fig.~\ref{fig:SFcurves}. In the bottom panel of Fig.~\ref{fig:sqcae2} we present examples of input and output snapshots obtained with CAE-2 for a helical and a skyrmion lattice configuration in the square lattice. It can be seen that CAE-2 is more adapted to the triangular based lattices, since in the square lattice it ``elongates'' skyrmions, and therefore the decoded images are significantly different to those presented in Fig.~\ref{fig:sqsnaps}. 

In conclusion,  a deeper CAE would be more specialized in the specific skyrmion lattices used for training. On the one hand, this could be desirable to distinguish other configurations from the RMSE, specially with the intention to detect exotic orderings. On the other hand, in this case, compared to the first CAE, it implied a poorer decodification for skyrmions in other geometries (the square lattice), and even for skyrmions in triangular lattices, which may not match exactly those constructed from the analytical parametrisation (Eq.(\ref{eq:skxpara})) used for training and validation of the algorithms, reducing the difference in the RMSE between the skyrmion and ferromagnetic phases,  which is the key in the anomaly detection techinique.

\newpage
%\bibliographystyle{apsrev4-2}
%\bibliography{MLproject}

%\bibliographystyle{apsrev4-2}
%\bibliography{MLproject}


\end{document}